# Towards Developing Brain-Computer Interfaces for People with Multiple Sclerosis

John S. Russo[1,], Tim Mahoney[1], Kirill Kokorin[1,2], Ashley Reynolds[1,4], Chin-Hsuan Sophie Lin[5*], Sam E. John[1,2*], David B. Grayden[1,2,3*]

1. Department of Biomedical Engineering, The University of Melbourne, Melbourne, Australia
2. Graeme Clark Institute, The University of Melbourne, Melbourne, Australia
3. Department of Medicine, St Vincent's Hospital, The University of Melbourne, Melbourne, Australia
4. Department of Neurosciences, St. Vincent's Hospital, The University of Melbourne, Melbourne, Australia
5. Melbourne School of Psychological Sciences, The University of Melbourne, Melbourne, Australia.

[*] Joint Senior authors

## Abstract

Multiple Sclerosis (MS) is a severely disabling condition that leads to various neurological symptoms. A Brain-Computer Interface (BCI) may substitute some lost function; however, there is a lack of BCI research in people with MS. To progress this research area effectively and efficiently, we aimed to evaluate user needs and assess the feasibility and user-centric requirements of a BCI for people with MS. We conducted an online survey of 34 people with MS to qualitatively assess user preferences and establish the initial steps of user-centred design. The survey aimed to understand their interest and preferences in BCI and bionic applications. We demonstrated widespread interest for BCI applications in all stages of MS, with a preference for a non-invasive (n = 12) or minimally invasive (n = 15) BCI over carer assistance (n = 6). Qualitative assessment indicated that this preference was not influenced by level of independence. Additionally, strong interest was noted in bionic technology for sensory and autonomic functions. Considering the potential to enhance independence and quality of life for people living with MS, the results emphasise the importance of user-centred design for future advancement of BCIs that account for the unique pathological changes associated with MS.

## 1 Introduction

People with Multiple Sclerosis (MS) may enhance their quality of life through Brain-Computer Interface (BCI) devices, a technology that allows an individual to control a virtual or physical device using their brain activity. However, the user requirements of BCI technology for people with MS are unknown.

MS is a chronic disease where the immune system mistakenly attacks the protective myelin covering of nerve fibres in the central nervous system, causing inflammation and damage (1). This leads to a range of highly variable symptoms, including muscle weakness, difficulty walking, impaired vision, and fatigue (2,3). Due to the variable progression of MS, people with MS may only experience mild symptoms or enter remission, while others experience disease relapses (flairs) and/or gradual accumulation of permanent disability, and premature death (4). The economic burden associated with the current treatments for MS, including pharmacotherapies, physical therapy, speech and language therapy, and occupational therapy (5,6), is approximately $6600 USD per person annually (7). This, combined with an average of 22 years of life lost per person and 1.2M disability-adjusted life-years across the population in the USA (2016) (4), highlights a severe reduction in quality of life for those with MS.

A BCI may provide an alternate and/or complementary means to restore independence for people with MS who experience temporary or permanent disability. BCIs could achieve this by assisting with every day, at-home activities, as has been shown for those with Amyotrophic Lateral Sclerosis (ALS) (8). In people with MS, investigation of BCI usefulness has been conducted in two previous studies demonstrating the potential for improved autonomy (9) and hybrid control of assistive technology during fatigue (10). An additional study suggested BCIs could be incorporated into a rehabilitation program using neuromuscular electrical stimulation (11). Furthermore, two studies provided some evidence for efficacy of visually controlled BCI in an individual with MS who was part of a larger heterogeneous group of paralysed people (12,13). BCIs may also have the potential to alleviate fatigue (14), which is one of the most common symptoms experienced by people with MS (15,16). The feasibility of decoding imagined movements was investigated in an individual with MS and demonstrated similar classification accuracy to neurotypical controls (17). This was expanded in a preliminary study using source localisation that aimed to provide a robust method for the temporal and spatially dynamic pathogenic processes of MS (18).

The small number of studies (9–13,17,18) highlight that BCI development for people with MS is relatively new. Additionally, these studies (9–13,17,18) did not describe the user requirements of BCIs for those with MS. To date, surveys investigating BCI user preferences have not explored the population with MS to a useful extent, with only brief mention in the International BCI Meeting series (19). Instead, surveys have focussed on populations with ALS (20) and spinal cord injury (21). The need for investigation in the population with MS was further emphasised in a clinician awareness survey (22). The survey (22) did not initially include MS as a patient population, but feedback from clinicians indicated people with MS may benefit from BCI implementation (22).

To progress this research area effectively and efficiently, a user-centred design strategy (23–26) that includes input from the potential user group is needed. This will enable development of technology that is suited to the users' needs. This approach marks a departure from conventional BCI research, wherein only a fraction (approximately 10% between 2015-2019) of studies tested BCIs with end users (27). Such a low ratio suggests that the user-centred design is minimally adopted in BCI research, thus resulting in insufficient investigation of how improvements could directly benefit end users within their own homes. This highlights a gap in the literature between researchers and end users, which this study aims to fill for the MS population.

As the first step in the user-centred design process, we explored the user requirements of people with MS (24). Our survey aimed to understand what BCI and bionic applications people with MS are most interested in and what kind of devices they prefer in the context of their symptoms and assistance needs. We hypothesised that (i) people with MS would be interested in adopting BCI technology, and (ii) those with reduced independence would prefer a higher-performing invasive BCI.

## 2 Methods

### 2.1 Survey

An online survey of people with MS was conducted to determine user preferences regarding BCI designs. Each respondent was asked to report their symptoms, what assistive technology they use, and if they require a carer or have difficulty using technology.

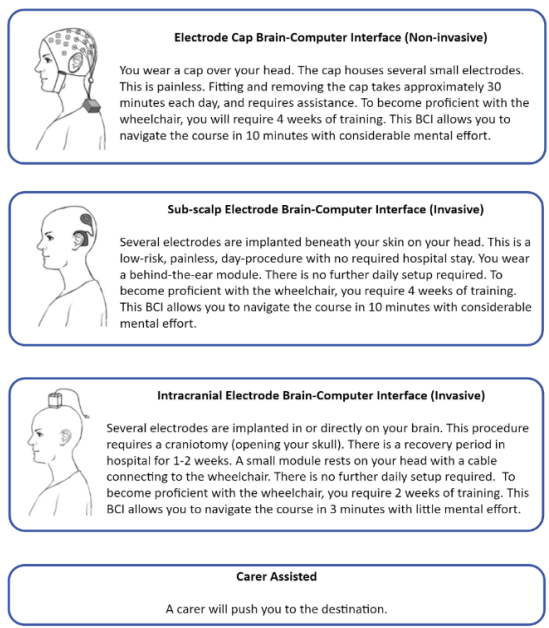

**Figure 1.** Ranking of BCI design options for the wheelchair scenario. Invasiveness, setup, training required, and baseline performance were described for each design option.

The respondents were then presented with three BCI applications: communication, wheelchair control, and robotic arm control, and asked to choose which application they believed would be most useful for them. For the application that they selected, the respondent was then presented with three BCI design options with different levels of invasiveness. Each option was described in terms of the associated hospital procedure, setup time, training time, and expected performance (Figure 1). Expected performance and training times were based on prior studies that implemented similar BCI systems (28–38), summarised in Table 1. As an additional fourth option, the respondent could choose to be assisted by a carer. After considering all provided information, the respondent was asked to rank each design option in order of preference.

To end the survey, respondents were asked to consider more broadly what aspects of their MS they would most like assistance with. Respondents could again select the BCI application they chose previously or an alternative bionic technology that could assist with functions such as temperature regulation or bowel and bladder control. For the full survey, refer to Supplementary Material.

### 2.2 Respondents

Respondents were recruited through survey advertisements posted on websites for people with MS, social media, and hospitals and clinics around Australia between 28 May – 1 November 2023. Inclusion criteria were people with MS over the age of 18 years. This study was approved by the University of Melbourne Human Research Ethics Committee (ID 26078). By completing the survey, respondents gave implied consent to participate in the study.

### 2.3 Analysis

Survey responses were qualitatively analysed for relationships between MS type and desired BCI functionality and outcome. Analysis of aspects of MS included time since diagnosis, type of MS, symptomology, carer reliance, and difficulties with using a computer or telephone. These responses were compared with desired BCI function and recording modality. The level of performance and invasiveness for each recording modality were compared with the time since diagnosis, type of MS, and symptoms. Selection of an ideal bionic device function was compared with the selected BCI function and used to highlight additional research directions that should be considered to better address the needs of people with MS.

|  | **Scalp-EEG** | **Sub-Scalp EEG** | **Intracranial** |
|---|---|---|---|
| **Invasiveness** | - Cap over the head<br>- Non-invasive<br>- Painless | - Module behind the ear<br>- Low risk<br>- Painless<br>- Day-procedure (no hospital stay) | - Small module rests on head with a connecting cable<br>- Implantation of electrodes<br>- Requires craniotomy, recovery for 1-2 weeks |
| **Setup** | - 30 minutes each day<br>- Requires assistance | No set-up post procedure | No set-up post procedure |
| **Training** | 4 weeks (36) | 4 weeks (36) | 2 weeks (38) |
| **Wheelchair performance** | Navigation in 10 minutes with considerable effort (28,29) | Navigation in 10 minutes with considerable effort (28,29) | Navigation in 3 minutes with little mental effort (37) |
| **Communication performance** | Type sentence in 1 minute with considerable effort (34) | Type sentence in 1 minute with considerable effort (34) | Type sentence in 15 seconds with little effort (35) |
| **Robotic arm performance** | Complete drinking task in 3 minutes with considerable mental effort (28) | Complete drinking task in 3 minutes with considerable mental effort (28) | Complete drinking task in 1 minute with little mental effort (30) |

**Table 1.** Critical parameters for each BCI scenario.

## 3 Results

### 3.1 Respondent Demographics

Survey responses were recorded from 34 adults with MS: 25 respondents (73%) were diagnosed with relapsing-remitting MS (RRMS) and 9 (26%) with primary or secondary progressive MS (S/PPMS). Most respondents (68%, n=23) were aged between 30 and 49 years (Figure 2a) and were from either USA (50%, n=17) or Australia (44%, n=15) (Figure 2b). Approximately half of the respondents (44%) had completed a postgraduate degree as their highest level of education (n=15) (Figure 2c).

### 3.2 Diagnosis and Symptomology

The majority (53%, n=18) of respondents were diagnosed with MS between 1-10 years ago, only six (18%) were newly diagnosed with MS (<1 year) and ten (29%) had MS for more than 10 years (Figure 3a). Of the symptoms reported (Figure 3b), fatigue (82%, n=28) and sensory symptoms (79%, n=27) were the most common. Looking at symptoms that could directly influence the need for a communication, robotic arm, or wheelchair controlled BCI, only six (18%) respondents reported a speech impairment, while 18 (53%) and 23 (68%) reported motor symptoms of the arms and legs, respectively. Other common symptoms were pain (62%, n=21), bladder/bowel dysfunction (50%, n=17), and depression (47%, n=16).

### 3.3 Assistance Needs

Respondents reported never (41%, n=14), sometimes (47%, n=16), or often (12%, n=4) requiring the assistance of a carer for everyday activities (Figure 4a), where a carer could be a professional, partner, family member, or friend.

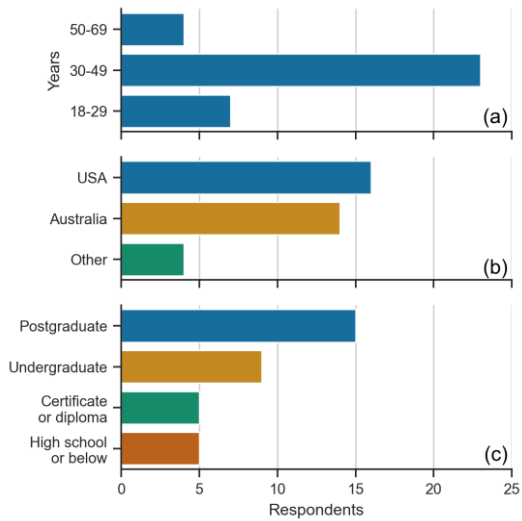

**Figure 2.** Respondent demographics. (a) Age. (b) Country of residence. (c) Highest level of education.

Most respondents (59%, n=20) never found using a computer or phone difficult, while 13 respondents (38%) found it difficult on some occasions (Figure 4b). The one (3%) respondent who often found using a computer/phone difficult only sometimes required the assistance of a carer. They were relatively older (50-69 years), had longstanding (21-30 years) progressive MS, and reported cognitive impairment.

The use of assistive technology was primarily focused on mobility (Figure 4c), which included walking aids (41%, n=14), wheelchair use (15%, n=5), and leg orthotics (18%, n=6). Some respondents used speech-to-text (21%, n=7) or text-to-speech (9%, n=3) software to assist with communication. In addition to the listed technologies, one respondent (3%) used grab rails and reaching aids.

The free-text answers describing what activities respondents found difficult or envisioned to find difficult in their future were combined into broad categories (Figure 4d) of working, self-care, and moving. MS could affect all these aspects of everyday life as highlighted by one respondent with multiple symptoms who stated that "everyday life [could] be difficult" and another who found "life in general" difficult.

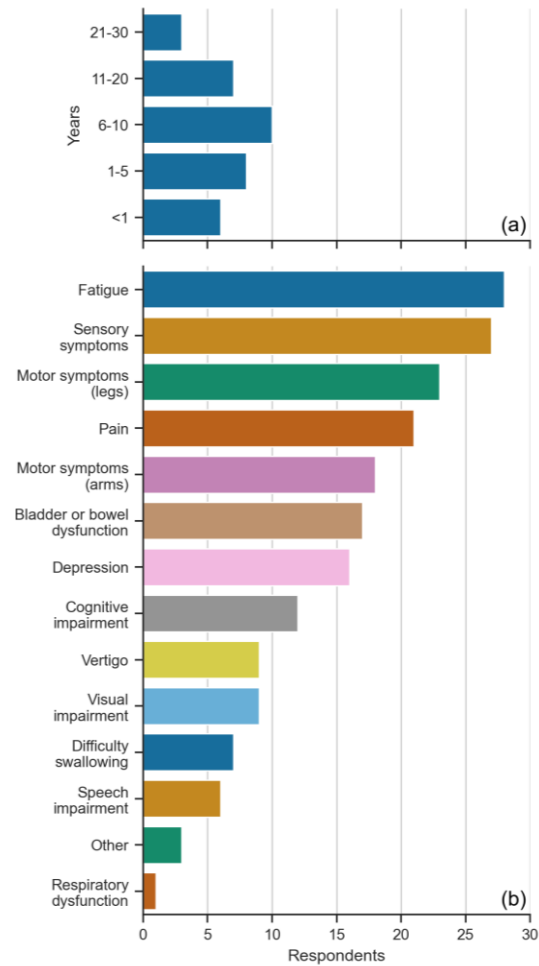

**Figure 3.** Diagnosis and symptomology. (a) Years since MS diagnosis. (b) Symptoms experienced organised by number of responses.

Respondents reported various aspects of self-care problematic (24%, n=8). As described by one respondent "bathing and dressing, as well as household chores like cooking and washing dishes were all difficult". They also found work difficult (18%, n=6). Working with a computer, using a mouse or keyboard, meetings, and lengthy periods of concentration were challenging and could cause nausea and fatigue. Moving their body was another challenging task (24%, n=8) for respondents, where walking even short distances and "standing for more than ten minutes" was difficult. One runner said: "on bad days I cannot run … I fall [down] a lot". Looking into the future, most respondents (44%, n=15) were worried that their ability to move, for example, "walking and using [their] hands", would deteriorate.

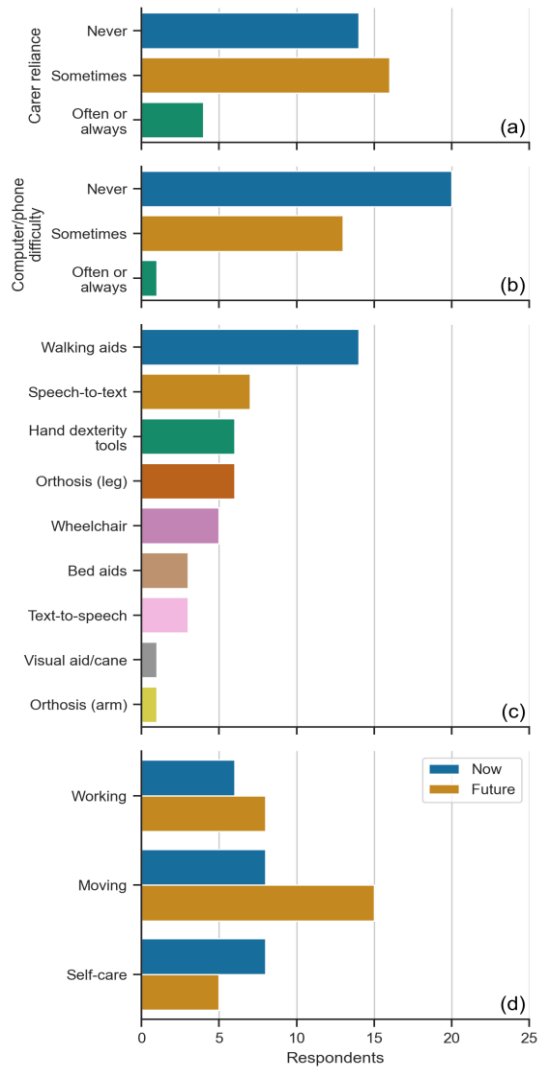

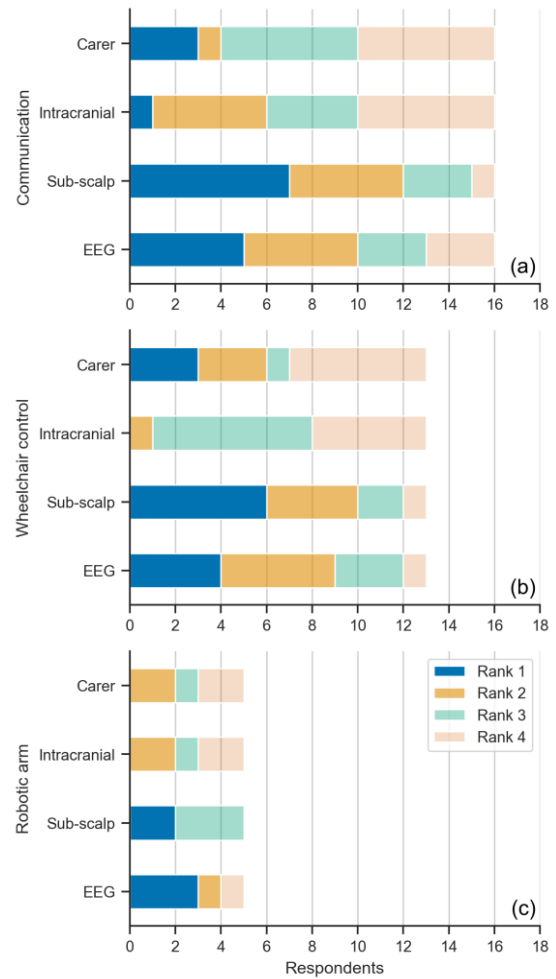

**Figure 4.** Assistance needs of respondents. (a) Require assistance of a carer. (b) Find using a phone or computer difficult. (c) Type of assistive technology used. (d) Activities they find difficult now or expect to find difficult in the future.

## 3.4 Desired Technologies

Of the three BCI scenarios surveyed, communication (47%, n=16) was the scenario chosen by the most respondents, followed by wheelchair (38%, n=13) and robotic arm control (15%, n=5) (Figure 5). Across all three scenarios, more respondents preferred the less invasive options of scalp (n=12) or sub-scalp electroencephalography (EEG) (n=15), compared to intracranial electrodes. Overall, 75%, 77% and 100% of respondents that chose the communication, wheelchair control and robotic arm control scenarios, respectively, preferred sub-scalp or scalp EEG.

**Figure 5.** Invasiveness preferences for BCIs. (a) Communication (n=16). (b) Wheelchair control (n=13). (c) Robotic arm control (n=5).

When asked about a selection of other bionic technologies, communication (6%, n=2), wheelchair control (12%, n=4), and robotic arm control (6%, n=2) were voted as less important than temperature regulation (18%, n=6) and bowel control (15%, n=5), although the differences between the responses were small (Figure 6). In a comparison of these results to the BCI scenario choices (Figure 5), most people who chose communication (41%, n=14), wheelchair (26%, n=9), or robotic arm control (9%, n=3), selected an alternative (non-BCI) technology. In addition to sensory symptoms, most people wanted a device for temperature regulation despite experiencing a broad range of other symptoms. However, respondents who currently experienced bowel/bladder dysfunction instead preferred a device for bowel control (Figure 7).

## 3.5 Symptoms vs. Scenario Selection

The distribution of symptoms factored by BCI scenario were compared to the overall distribution of symptoms within the entire study population (Figure 8). The distribution of the most prevalent four symptoms (fatigue, sensory symptoms, motor symptoms in the arms, and pain) between BCI scenarios was similar. As expected, respondents who experienced motor symptoms in the arms favoured the robotic arm scenario, and those that experienced speech impairment preferred a BCI for communication. A BCI for communication was also more frequently favoured by those reporting cognitive impairment, respiratory dysfunction, and depression, which can all impair the ability of a person to communicate. Respondents with bowel and bladder dysfunction preferred a BCI for wheelchair control or communication, with none selecting the robotic arm scenario. A BCI for robotic arm control was more commonly selected if respondents experienced fatigue, sensory symptoms, pain, vertigo, visual impairment, difficulty swallowing, and other symptoms, such as nausea (free text response labelled as 'Other' in Figure 8).

As depression can affect all aspects of daily living, it is essential to understand which MS symptoms often occurred together with depression to understand how depression affects BCI scenario selection. Respondents who reported depression (n=16) more frequently reported symptoms of fatigue (n=15), sensory disturbances (n=13), motor symptoms of the legs (n=10), cognitive impairment (n=9), pain (n=9), bowel or bladder dysfunction (n=8), and motor symptom of the arms (n=7). Only five respondents with depression reported difficulty with speech. Despite these associated symptoms, people with depression and MS ranked a BCI for communication the highest.

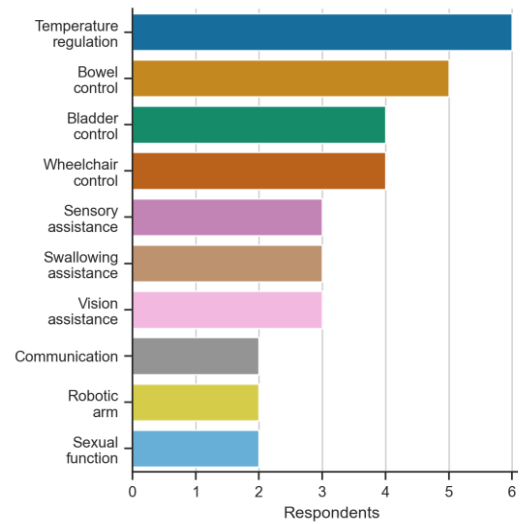

**Figure 6.** Preferred bionic technology.

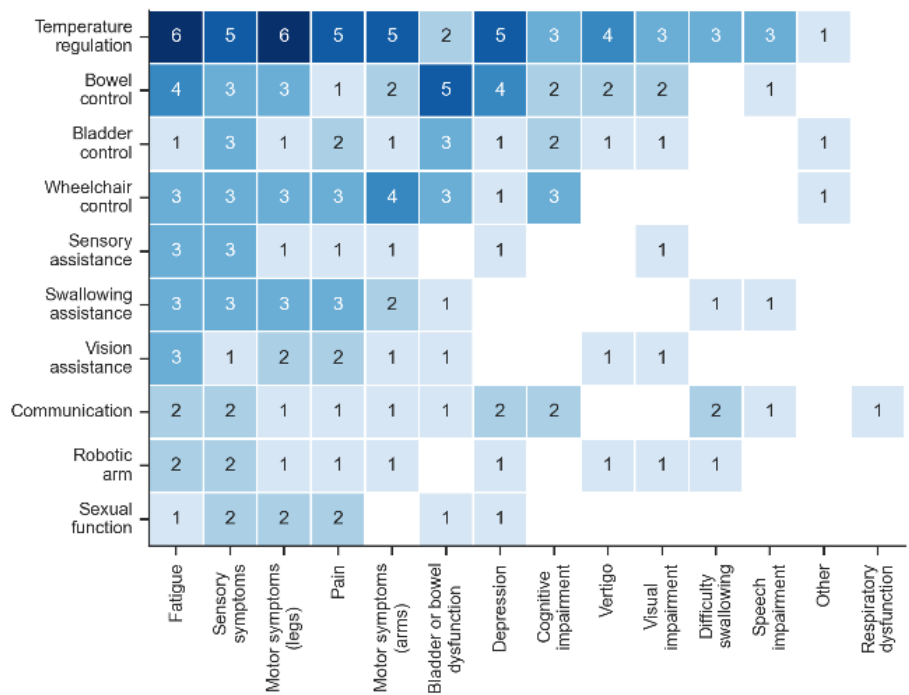

**Figure 7.** Number of respondents with a given symptom who selected a specific technology.
Page **7** of **18**

## 3.6 Assistance Needs vs. Desired Technology

The level of independence of each respondent, reflected by their reliance on a carer and difficulty using a phone or computer, appeared to have little effect on preferred device type (Figure 9). Of the four respondents who required a carer often, most (n=3) preferred the scalp EEG system (followed by sub-scalp, n=1), selecting the less invasive device options over intracranial BCI or continued carer assistance. Similarly, most of those who sometimes required carer assistance preferred BCI solutions (n=9, sub-scalp; n=4, EEG; n=1, intracranial) over a carer assisted solution (n=2). The one respondent who reported often having difficulty using a phone or computer also opted for a sub-scalp BCI as their top preference. The one respondent who opted for an intracranial device reported assistance by a carer was sometimes required, and never having difficulty with a phone or computer.

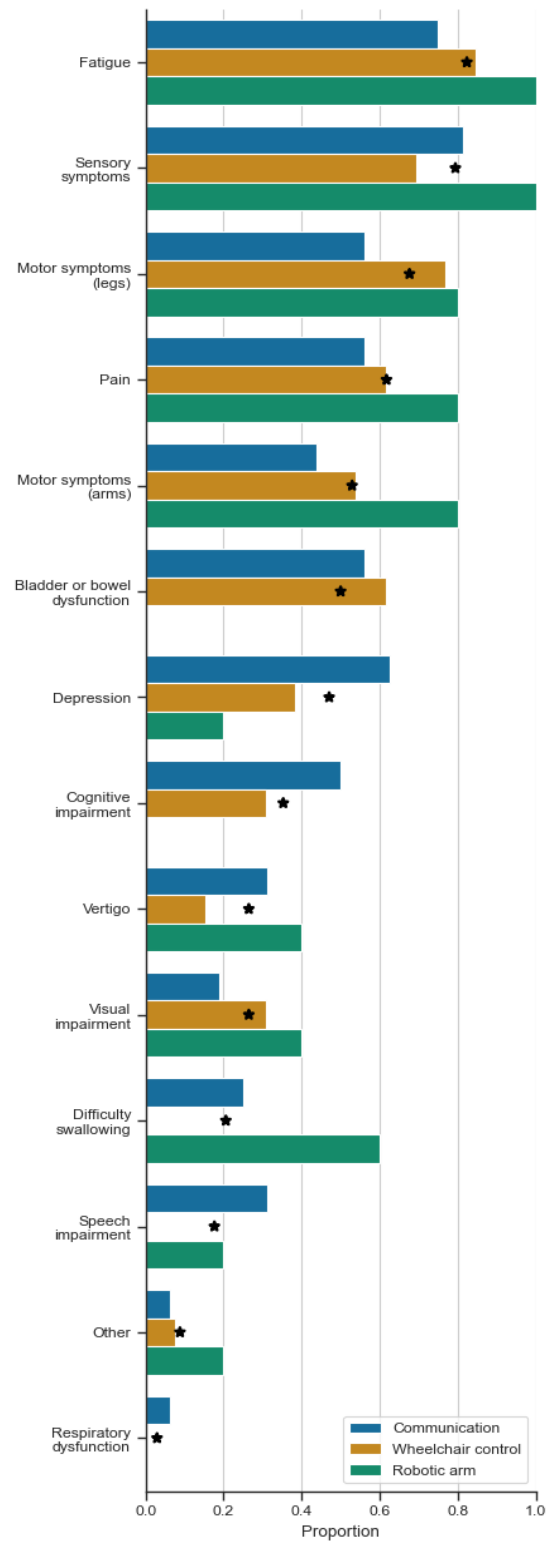

**Figure 8.** Proportion between 0 (0%) and 1 (100%) of respondents who had a specific symptom out of the total number who chose each BCI scenario. The stars indicate the proportion of all respondents who had each symptom.

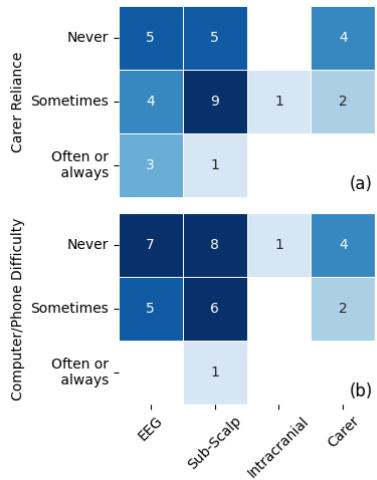

**Figure 9.** Comparisons between the highest ranked BCI device type and (a) how often a respondent requires the assistance of a carer or (b) finds using a computer and/or phone difficult. The numbers indicate how many respondents are in each category.

## 4 Discussion

### 4.1 Overview

This study is the first to identify and report user expectations of BCI systems from people with MS. We established the necessary insights to guide future research goals in BCI and bionic technologies for this distinct condition. Our results show that people with MS are interested in using BCIs and were more receptive to a non-invasive or minimally invasive BCI as the optimal device (Figure 5) that balances the benefits (e.g. improved performance) against the associated risks associated with surgery and device aesthetics (summarised in Table 1). If they had to choose one of three possible BCI functions, a BCI to support communication as opposed to moving a wheelchair or robotic arm was preferred (Figure 5). However, people ideally wanted bionic technologies to manage abnormal sensory symptoms and functions of the autonomic nervous system as opposed to the traditional BCI functions previously mentioned (Figure 6). Furthermore, the preference for a non-invasive or minimally invasive BCI did not appear to be influenced by current assistance needs (Figure 9), challenging the hypothesis that reduced independence correlates with a tendency to select higher-performing invasive BCIs. Invasiveness Preferences

Most people (>75%) preferred sub-scalp or scalp EEG over invasive BCIs that needed brain surgery. It must however be considered that in a survey it may be difficult to adequately capture the tedious setup time of EEG and increased performance of intracranial BCI. There was a slight preference for sub-scalp (n=15) over scalp EEG (n=12). This suggests respondents perceive that a sub-scalp EEG may provide an appropriate balance between invasiveness and performance for various degrees of independence. Long-term sub-scalp EEG is a recent technology, currently used for epilepsy monitoring ((39) for a review). These systems provide a minimally invasive alternative to standard EEG without the need for ongoing donning and doffing procedures. To the best of our knowledge, this survey is the first to assess potential user preferences of sub-scalp EEG for a BCI application. The results encourage further investigation that assesses sub-scalp EEG signal quality and BCI performance, as well as preferences of other potential BCI user cohorts.

Only one respondent preferred intracranial electrodes despite this design option providing the user with the highest performance. Similarly, previous work found persons with spinal cord injury were more likely to adopt non-invasive BCI design options (40). This result suggests aversion to intracranial BCIs, which may be due to invasiveness, in-hospital recovery, or aesthetics, outweighs the benefits of increased BCI control. Unfortunately, respondents were not given the opportunity to express the reasoning behind their rankings. As such, it is unclear what aspects of the design options were most important. This could be explored in subsequent work to ensure BCI designs meet user expectations.

## 4.2 Providing Assistance with Brain-Computer Interfaces

The current assistance needs of respondents, in terms of support from a carer and difficulty using technology, did not appear to directly influence BCI preferences, with most preferring a scalp or sub-scalp EEG system (Figure 9).

The one respondent who often found a computer/phone difficult to use selected a sub-scalp EEG, illustrating that people who are less proficient with technology are still interested in BCI. However, this is not certain given our survey did not capture the reason people found using technology difficult. Furthermore, for all scenarios, most respondents who were assisted by a carer preferred a non/minimally invasive BCI solution over carer assistance. This suggests that people with MS who has experienced the benefits that a carer provides are still interested in using a BCI for assistance.

## 4.3 BCI for Communication for People with MS

Our findings indicate people with MS prioritise a BCI system that can support communication. However, the impact of MS on communication is multifaceted (41). Symptoms such as dysarthria, aphasia, hearing impairment, problems with vision, cognitive impairment, and fatigue hinder the ability to effectively communicate and opportunities to communicate (41). Additionally, reduced hand dexterity and somatosensory disturbances can pose challenges in written, electronic, and telecommunication means of communication. These difficulties are especially important in young, working-age adults, a demographic commonly affected by MS (41).

Moreover, depression affects communication through psychological barriers (e.g., minimising thoughts and feelings, negative expectations of interactions, and feeling too overwhelmed to communicate (42,43)) as opposed to physical barriers, and it is estimated that one in two people with MS will experience major depression (44). Conversely, impaired communication can lead to feelings of isolation, and both factors are known risks for depression. Fortunately, the use of communication technologies can mitigate these risks (45). Therefore, it is important to draw attention to our results, which support the possibility that people might want a BCI to overcome the psychological barriers of communication, since respondents with depression preferred a BCI system to aid communication over other functions. This is despite only a minority of this sub-cohort reporting current or anticipated future difficulty with speech.

Given our survey was only designed to broadly gauge user expectations of BCIs, we barely scraped the surface of how BCIs can aid communication in this group. Therefore, future research aimed at developing this technology would benefit from collaboration with MS specialist speech pathologists to determine which components of communication (e.g., speech/text production, aiding word finding, and cortical visual/auditory restoration) would benefit from BCI support. This will be essential to aligning BCI functionality with various expectations of the end-user.

## 4.4 Adapting to a Dynamic Condition

A successful BCI for MS requires the ability to adapt to long-term, medium-term, and short-term changes to the signal that is used to decode user intention. We define long-term changes as those caused by the progressive neurodegenerative processes that occur over years. Medium-term refers to MS flairs where new regions of the central nervous system are affected; these new symptoms may only be present for days to months if treated with an effective intervention. Short-term refers to the most common symptom, fatigue, which affects brain function over hours and days.

As BCIs were initially designed for people with locked-in syndrome, typically due to ALS (27),

researchers already focus on adaptive technology to cope with the long-term changes in brain activity that BCI users may experience (23). However, the challenge with MS stems from the nature of the disease progression, which is more unpredictable and variable, both between and within individuals and over both time and location (2). This implies that the EEG signal may change temporarily or permanently due to inflammatory processes, demyelination, axonal and cortical loss, and immunomodulating therapy. As both structural and functional cortical changes occur with MS (5,46), the BCI system could be implanted in the target cortical region that also has the lowest probability of developing MS lesions and permanent tissue loss (i.e., avoiding sulci and areas of low cerebral artery perfusion) (46). A BCI may also be made more robust to medium- and long-term changes in signal quality by decoding source rather than sensor activity. The initial feasibility of extracting source-level features for decoding in those with MS has been shown (18).

Similar to previous studies (15,16), our results showed that fatigue is one of the most common symptoms experienced by people with MS. This symptom is not unique to MS and presents a key challenge for any BCI technology where fatigue can negatively affect overall system performance (47). For an EEG BCI, this drop in performance may even be a result of fatigue directly modulating the patterns of brain activity (47,48), thus making it harder for the system to decode the intent of the user.

Overall, it is critical that a BCI can operate while a person is fatigued, as that is when they are likely to need the system the most. In this situation, a BCI can be designed to directly monitor the mental state of the user and adapt its level of assistance (49). However, this requires being able to detect what mental state the user is in (either based on their brain activity (14,50,51) or control behaviours (49)), thus making the system more complex and requiring additional training data. A simple alternative would be to allow the user instead to select manually the level of assistance and type of BCI control (e.g., switching from a virtual keyboard to yes/no communication when fatigued).

Even without trying to adapt to the mental state of the user, current BCIs require the collection of training data for the creation of a custom machine-learning model that decodes brain activity. Continuously gathering training data and re-calibrating the model improves decoding performance (38) and may address the challenges associated with changes in brain activity over time. However, the need for recalibration could prevent someone from immediately using their BCI and burden the user, especially if they are prone to fatigue. To address this problem, BCI training should be infrequent but engaging, with decoders designed to be trained on small, high-quality datasets (52). In the ideal case, once a user has calibrated the system once, it could continue to automatically adapt based on everyday use, without dedicated training sessions (53).

## 4.5 Bionic Technologies to Address Symptoms of Multiple Sclerosis

Our results suggest that bionic devices that, in the future, could manage sensory or autonomic symptoms (e.g., through somatosensory cortical stimulation (54) or regulate bowel and bladder function (55) may be as relevant to people with MS as typical active BCI applications. Bowel/bladder control and temperature regulation appeared to directly influence the technology of choice (Figure 7). This is unsurprising given bowel and bladder dysfunction are strongly correlated with a poor quality of life (56). Furthermore, abnormal temperature regulation combined with heat sensitivity, impairs conduction of demyelinated nerves. This transiently worsens all symptoms including fatigue, and there are limited treatment strategies to address this problem (57).

Our findings are likely to reflect the preferences of the broader MS population, given the symptoms experienced by our respondents were consistent with larger cohort studies of MS (15,16). Moreover, as MS can cause lesions in the spinal cord, it is expected given our findings also align with preferences of people with spinal cord injury (58).

Due to the scope of our survey, we grouped the multitude of sensory symptoms experienced by people with MS (e.g., paraesthesia, loss of sensation, hyperesthesia, heat sensitivity, L'Hermitte's sign, pain, impaired proprioception, and visual disturbances (2)) into the coarse categories of pain, vertigo, vision impairment, and sensory symptoms. Respondents may also have provided answers based on their perceived future symptoms when selecting a technology and requires clarification. To better understand how bionic technologies can best assist people with MS, we recommend that future work explore user preferences for bionic devices that address specific sensory or autonomic symptoms.

### 4.6 Limitations

Our survey did not capture how people in a low-resource setting may evaluate the technologies as all our respondents were from high-income countries, which have higher rates of MS (4). Additionally, many of the respondents had a postgraduate education, which may have affected their willingness to adopt technology.

The use of an online survey likely biased our sample towards people who could easily use a computer or phone, while recruitment from MS clinics would lead to respondents with a reduced level of disability compared with those who disengaged from healthcare services (59). Consequently, our sample only had a few respondents who often needed the assistance of a carer or found the use of a computer/phone difficult. Therefore, we remain cautious in our conclusion that people with more severe symptoms are less open to invasive BCIs. Future work is required to understand the bionic technology preferences of this user demographic, especially since these technologies could deliver the most benefit to this cohort.

The perspectives of carers, especially informal carers (such as family or friends), were not considered in our study. As the disease and disability associated with MS progresses, it is common for partners of individuals with MS to assume the role of an untrained and unpaid carer (60). It is important to consider that carer strain affects 42% of carers for people with MS (61). This leads to carers struggling to perform the caregiving role, reducing both persons' quality of lives, and increasing rates of neglect, abuse, morbidity, need for respite services, permanent institutionalisation and mortality (61). Assistive technologies like BCI have the potential to alleviate demands on carers, allowing them more time for paid employment, rest, and sleep. Reducing the burden of care with assistive technology may be an underexplored solution that could improve the long-term quality of life for the person with MS and their carer.

Adequately describing the experience of using a BCI is challenging, especially when communicating differences in performance. It is possible that our scenario descriptions did not adequately capture the frustration of using a lower information transfer rate system, thus biasing selection towards the non/minimally invasive BCIs. However, it is expected that most people in general would opt for a device associated with fewer risks related to implantation (40). Future work should collaborate with user experience designers to consider strategies for better explaining system performance, setup time, and training time that does not require participants to take part in a full BCI experiment. This could be done by including videos displaying system performance, interviews with people who have used the systems, and deploying systems that simulate the experience (e.g., an online keyboard with a given information transfer rate).

Furthermore, the design options presented to respondents were based on previous research, which may not represent the technologies that will be available in the near future. With considerable investment and clinical trials underway for both intracranial (62) and novel endovascular BCIs (63), combined with research efforts into artificial intelligence, future BCI devices should have improved performance, lower training commitments, and reduced associated risks. However, challenges to produce high performing BCIs may persist when translating these technologies into the home environment. As more BCIs become commercially available and better understood, regularly re-evaluating design options would improve our understanding of the user specifications of people with MS.

## 5 Conclusions

Multiple Sclerosis (MS) is a severely disabling condition that has received minimal consideration from Brain-Computer Interface (BCI) researchers. Our survey of people with MS indicated that we need to revise this viewpoint. We provided evidence that people at different stages of MS are interested in BCI technology and would opt to use a BCI over a carer, particularly if the device was non-invasive or minimally invasive. We further provided key research goals for this technology, including robust signal acquisition and novel training paradigms in anticipation of the dynamic temporal and spatial pathological changes associated with MS. Additionally, respondents expressed a strong interest in bionic technology for therapeutic applications, such as assistance with temperature regulation and bowel/bladder control. Considering the potential to enhance independence and quality of life for people living with MS, we advocate for advancement in this research field, emphasising a user-centred design approach as its cornerstone.

## 6 Declarations

**Corresponding author**

John S. Russo: russoj1@student.unimelb.edu.au ORCID 0000-0003-3010-8541

**Co-author details**

Tim Mahoney: tbmahoney@student.unimelb.edu.au ORCID 0000-0002-2147-7698

Kirill Kokorin: kkokorin@student.unimelb.edu.au ORCID 0000-0002-9003-6917

Ashley Reynolds: ashleyr@student.unimelb.edu.au ORCID 0000-0003-3614-5300

Chin-Hsuan Sophie Lin: chinhsuan.lin@unimelb.edu.au ORCID 0000-0002-2156-1830

Sam E. John: sam.john@unimelb.edu.au ORCID 0000-0003-3780-2210

David B. Grayden: grayden@unimelb.edu.au ORCID 0000-0002-5497-7234

**Funding**

A.R., T.M., and J.R. receive funding from the Australian Government Research Training Program Scholarship from the University of Melbourne. K.K., S.E.J. and D.B.G. are supported by the ARC (Australian Research Council) Industrial Transformation Training Centre in Cognitive Computing for Medical Technologies (IC170100030). T.M. is supported by the Elizabeth and Vernon Puzey Scholarship. C.S.L is supported by the Melbourne School of Psychological Sciences.

**Disclosure statement**

The authors report there are no competing interests to declare.

**Availability of data and materials**

The datasets generated and/or analysed during the current study are not publicly available because they contain sensitive personal information but are available from the corresponding author on reasonable request.

**Ethics approval and consent to participate**

This study was approved by the University of Melbourne Human Research Ethics Committee (ID 26078). Respondents gave implied consent to take part in the study through completing the survey. The research was conducted in accordance with the principles embodied in the Declaration of Helsinki and in accordance with local statutory requirements.

**Acknowledgements**

We would like to express our sincere gratitude to Ms. JingYang Liu for her valuable contributions to this research; she played a crucial role in creating the illustrations for the survey.